\documentclass{article}

\usepackage{graphics}
\usepackage{units}
\usepackage{amsmath}
\usepackage{amssymb}
\usepackage{graphicx}
\usepackage{wasysym}
\usepackage{esint}
\usepackage{geometry}
\begin{document}
\title{Thermal optimization of Curzon-Ahlborn heat engines operating under 
some generalized efficient power regimes}

\author{S. Levario--Medina\textsuperscript{a}, G. Valencia--Ortega\textsuperscript{b} and L. A. Arias--Hernandez\textsuperscript{c}\\
Departamento de F\'{i}sica, Escuela Superior de F\'{i}sica y Matem\'{a}ticas, Instituto\\
Polit\'{e}cnico Nacional, U. P. Zacatenco, edif. \#9, 2o Piso, Ciudad de M\'{e}xico,\\
07738, M\'{e}xico, levario@esfm.ipn.mx\textsuperscript{a}; gvalencia@esfm.ipn.mx\textsuperscript{b} and\\ larias@esfm.ipn.mx; 
http://orcid.org/0000-0003-4054-5446\textsuperscript{c}}
 
\maketitle

\begin{abstract}
In order to establish better performance compromises between the process functionals of a heat engine, in the context of finite–time thermodynamics (FTT), we propose some generalizations for the well known Efficient Power function through certain variables called ''Generalization Parameters''.
These generalization proposals show advantages in the
characterization of operation modes for an endoreversible heat engine model.
In particular, with introduce the $k$--Efficient Power regime. For this obejctive function we find the performance of the 
operation of some power plants through the parameter $k$. Likewise, 
for plants that operate in a low efficiency zone, within a configuration space, 
the $k$ parameter allow us to generate conditions for these plants to operate inside of a high efficiency and low dissipation zone.

\vspace{0.3cm}05.70-Ln Nonequilibrium and irreversible thermodynamics; 84.60.Bk Performance characteristics of energy conversion system; figure of merit
 89.30.-g Fossils fuels and nuclear power.

\end{abstract}
%


%
\section{Introduction}
\label{intro}
One of the main objectives in the study of thermal engines is to improve their energy performance. Some efforts have focused on reducing the effects associated with the dissipation of energy during the operation of thermal engines \cite{BarrancoAngulo96,CurtoMedinaGuzmanAngulo11,SanchezMedinaCalvo13}, 
other ones have directed to design new energy converters that use higher energetic density fuels \cite{DeVos95,Miyazaki00,DharKumar14}. 
One example of the above is the development of different types of power plants, whose purpose has been to increase the efficiency of energy conversion. 
The result of this effort has led to the construction of combined cycle 
plants, which have exhibited a good improvement in their efficiencies, but their energy dissipation in form of heath to the
atmosphere continues to be wasted in large quantities. Nevertheless,
it is not yet achieved to a large extent establish a compromise between
the way of operating an energy converter and a modification in its
engineering, without having to build a whole new machine \cite{SilvaArias13,DeVos92,Arias11}. 
To establish this trade off, we use a  FTT description which has emulated the performance of 
an energy converter by using diverse objective functions  \cite{Bejan96,Bejan88,Hoffmann97,Hoffmann03,Durmayaz04,Wu99,Chen04}.
This new way of studying thermal machines led Curzon and Ahlborn to
propose in 1975 \cite{CurzonAhlborn} one of the first models that
presents the power output as an objective function. This function
distinguishes the maximum power output as a well--known mode of operation
for heat engines. Ever since, several authors have proposed other
objective functions, which are not only formed with the individual process functionals
(power output, efficiency and dissipation), but also present
a good compromise between them \cite{Angulo,AriasAngulo,Calvo}. All
above with the aim of characterizing optimal operating regimes. However,
despite the current approaches that are describing some of the energy
converters in good performance regimes, no systematic method has
been found with which those type of objective functions can be built.
In recent papers, it has been possible to find and
quantify the best compromise between the process variables involved through certain criteria of merit,
and they have given rise to some proposals for their generalization. In
particular, these happened for the ecological ($E=P-\Phi$, being $P$ the power output and $\Phi$ the dissipation function) \cite{Angulo,AriasAngulo,FunCom}
and omega  $(\Omega=E_{u,e}-E_{u,p})$ functions \cite{Calvo,lilian,LevarioArias}.
These new generalized functions contain the so-called generalization
parameters, for example for the case of generalized ecological function $E=P-\epsilon\Phi$ the generalized paremeter is $\epsilon$ and for the generalized omega function $\Omega=E_{u,e}-\lambda E_{u,p}$ is $\lambda$. 
Which in rule allows out obtaining a large family of objective functions \cite{LevarioArias}.
In the endoreversible, the non-endoreversible and irreversible models, the generalized ecological and omega functions 
have the same maxima \cite{SilvaArias13,Arias11,lilian,LevarioArias,Arias10}.

\begin{figure*}
\centering
\resizebox{0.5\textwidth}{!}{
\includegraphics[width=6cm]{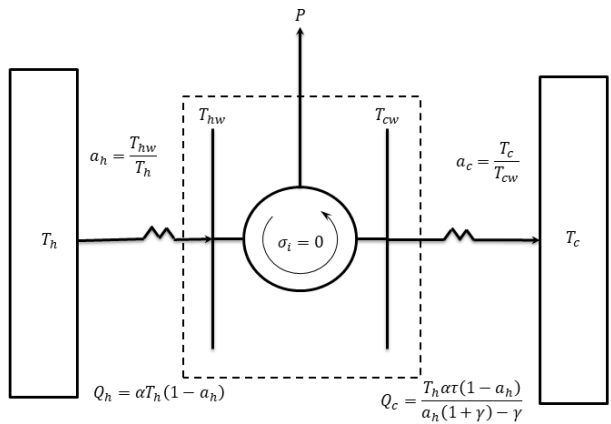}}
\caption{The well-known diagram of a Curzon--Ahlborn
(CA) heat engine.}
\label{fig:Motor-tipo-CAN}
\end{figure*}

Another proposal of commitment function was defined by Stucki \cite{Stucki} within the context of the Linear Nonequilibrium Thermodynamics called "economic power output", which describes with certain approximation, the
oxidative phosphorylation process that is involved in the synthesis
of adenosine triphosphate inside mitochondrias. After that, Yilmaz \cite{Yilmaz} in an independent way proposed the
same Stucki's objective function but in the frame of the FTT. His intention was to get a trade-off between
the delivered power and the efficiency for a heat engine. He called this
new function: "Efficient Power", defined as the product of the
power output by the efficiency:

\begin{equation}
P_{\eta}=P\eta.\label{eq:DefPotEfi}
\end{equation}

The general idea of all above mentioned works, in which are found
functions that accomplish good performance regimes for energy conversion processes,
is that the modes of operation are physically attainable. It leads
us to the freedom to propose new objective functions at best or set
up other generalizations for the functions already known, taking care
of the limits of its physical validity. Although the study carried out
in this work is effected for an endoreversible model and with linear
heat transfer laws, this model is able to set operating ranges for
more real models, focusing attention on the mechanisms of heat exchange
that in heat engines could be modified to what we have called
"restructuring conditions".

As we have pointed out, we are representing the irreversibilities
between the working substance and the external energy reservoirs through
a Newtonian heat transfer law. Therefore, we can write the input $\left(Q_{h}\right)$ and output $\left(Q_{c}\right)$ heat fluxes
as:

\begin{equation}
Q_{h}=\alpha\left(T_{h}-T_{hw}\right),\label{eq:C1LtN}
\end{equation}

\begin{equation}
Q_{c}=\beta\left(T_{cw}-T_{c}\right),\label{eq:C2LtN}
\end{equation}
where $\alpha$ and $\beta$ are the thermal conductances,
whereas that $T_{hw}$ and $T_{cw}$ are the temperatures of the working
substance. In addition, as in the Curzon--Ahlborn (CA) model,
the responsible processes in the production of work can be identified,
as the interchange of energy between the working substance and the reservoirs (see Fig. \ref{fig:Motor-tipo-CAN}). The total entropy production of the heat engine is:
\begin{equation}
\sigma_{T}=\sigma_{e}+\sigma_{ws}\geq0,\label{eq:EntroTS}
\end{equation}
with $\sigma_{e}$ is the entropy production at the coupling between the working substance and the surroundings and $\sigma_{ws}$ is the working substance entropy production. In particular, for the model shown in Fig.
\ref{fig:Motor-tipo-CAN}, $\sigma_{e}$ is given as:

\begin{equation}
\sigma_{e}=\frac{Q_{c}}{T_{c}}-\frac{Q_{h}}{T_{h}},\label{eq:Entros}
\end{equation}
while $\sigma_{ws}$ is:

\begin{equation}
\sigma_{ws}=\frac{Q_{h}}{T_{hw}}-\frac{Q_{c}}{T_{cw}}+\sigma_{i},\label{eq:EntoSt}
\end{equation}
being $\sigma_{i}$ the internal entropy production which is attributed
to different causes, such as turbulence, viscosity, among others. The working substance operates in cycles, then $\sigma_{ws}=0$ and for the endoreversible model, we have $\sigma_{i}=0$, so the Eq. \ref{eq:EntoSt} can be rewritten as follows,

\begin{equation}
\frac{Q_{h}}{T_{hw}}=\frac{Q_{c}}{T_{cw}}.\label{eq:HipEndoR}
\end{equation}
Eq. \ref{eq:HipEndoR} represents the well-known endoreversibility
hypothesis (Citar Rubin). While the surroundings entropy production continues to fulfill that,

\begin{equation}
\frac{Q_{c}}{T_{c}}-\frac{Q_{h}}{T_{h}}>0.\label{eq:EntroSisEndo}
\end{equation}

In the following, we express the heat fluxes $Q_{h}$ and $Q_{c}$
in terms of an auxiliary variable called high reduced temperature
$\left(a_{h}=T_{h}/T_{hw}\right)$, thus:

\begin{equation}
Q_{h}\left[\alpha,T_{h},a_{h}\right]=\alpha T_{h}\left(1-a_{h}\right)\label{eq:Q1Endo}
\end{equation}
and
\begin{equation}
Q_{c}\left[\alpha,\gamma,T_{h},\tau,a_{h}\right]=\frac{\alpha T_{h}\tau\left(1-a_{h}\right)}{a_{h}-\gamma\left(1-a_{h}\right)},\label{eq:Q2Endo}
\end{equation}
where $\gamma=\alpha/\beta$ and $\tau=T_{c}/T_{h}$. Besides, $Q_{c}$
can be expressed in terms of $a_{h}$ by Eq. \ref{eq:HipEndoR}
which shows the internal irreversibilities occurring in the energy conversion
process. In principle, the heat fluxes $Q_{h}$ and
$Q_{c}$ can characterize the behavior of this CA heat engine. Although,
the aforementioned process variables become the objective functions
of our interest.

The process variables represent the energetic behavior of an engine,
and under the considerations we have made in the paragraph after Eq. (\ref{eq:EntoSt}) for the heat fluxes $Q_{h}$
and $Q_{c}$, we have the following characteristic functions:

\begin{eqnarray}
P[\alpha,\gamma,T_{h},\tau,a_{h}] & = & Q_{h}-Q_{c}\nonumber \\
 & = & \frac{\alpha T_{h}\left(1-a_{h}\right)\left[a_{h}-\gamma\left(1-a_{h}\right)-\tau\right]}{a_{h}-\gamma\left(1-a_{h}\right)},\label{eq:PeMEndo}
\end{eqnarray}
\begin{eqnarray}
\eta\left[\gamma,\tau,a_{h}\right] & = & 1-\frac{Q_{c}}{Q_{h}}\nonumber \\
 & = & 1-\frac{\tau}{a_{h}-\gamma\left(1-a_{h}\right)}\label{eq:EfieMEndo}
\end{eqnarray}
and
\begin{eqnarray}
\Phi\left[\alpha,\gamma,T_{h},\tau,a_{h}\right] = T_{c}\sigma_T & = & Q_{c}-\tau Q_{h}\nonumber \\
 & = & \frac{\alpha T_{h}\tau\left(1+\gamma\right)\left(1-a_{h}\right)^{2}}{a_{h}-\gamma\left(1-a_{h}\right)}\label{eq:DisEMEndo}
\end{eqnarray}
In principle, these functions contain all the thermodynamic information
of the system, they are normally used to limit the operating ranges
of the real thermal machines. 

This article is organized as follows, in Sect.~\ref{sec:2}, we introduce three generalization proposals for the Efficient
Power $\left(P_{\eta}\right)$, and we make an energetic analysis
of each of them. In Sect.~\ref{sec:3}, we characterize the behavior of some
power plants by using the proposal of the $k$--Efficient Power,
since this generalized function identifies all the physically accessible points
of the Power Output \textit{vs.} Efficiency curve. Also, we show how the
generalization parameters allow us to modify aspects of the operation
of a heat engine. Finally, in Sect.~\ref{sec:4} we enunciate the concluding
remarks of this work.

\section{Generalizations of the Efficient Power}
\label{sec:2}
One of the problems in the study of energy conversion
for any heat engine, is characterize all the operation modes to which a thermal machine can gain access \cite{LevarioArias}, this problem  lies in the lack of objective functions associated to those modes. In this work, we propose a generalization of the objective function ``Efficient Power'', in order to describe the largest number of accessible operation modes for a heat engine \cite{Zhang17}. This generalized function can be written as,

\begin{equation}
P\eta_{g}=\left(P^{\sigma}\eta^{\upsilon}\right)^{\chi}.\label{eq:GenPotEfiG}
\end{equation}

However, this generalization is reduced to three particular cases,
which only lead to define a single generalization parameter, because
in \cite{LevarioArias} it has been shown that a single parameter
is enough to associate the particular performance of a thermal engine
with a specific mode of operation. In each of them, different modes
of operation can be achieved for the CA energy converter model.

When $\chi$ parameter is $\chi=\nicefrac{1}{\sigma}$, we can define
the exponent that acts in efficiency as $k=\nicefrac{\upsilon}{\sigma}$
and we get the commitment function named as $k$--Efficient Power \cite{LevarioArias}.
Given by:

\begin{equation}
P\eta_{k}=P\eta^{k}.\label{eq:kPotEfi}
\end{equation}

If $\chi=\nicefrac{1}{\upsilon}$, we can determine the exponent on
the power output as $l=\nicefrac{\sigma}{\upsilon}$. We call this
function $l$--Efficient Power, whose expression is,

\begin{equation}
P\eta_{l}=P^{l}\eta.\label{eq:lPotEfi}
\end{equation}

Finally, in case $\upsilon=\sigma=1$, we have the function $\chi$-Efficient
Power, defined by

\begin{equation}
P\eta_{\chi}=\left(P\eta\right)^{\chi}.\label{eq:mPotEfi}
\end{equation}

Upon this, we will analyze some of the characteristics that each of
the different generalization proposals has got.

\begin{figure*}
\centering
\resizebox{0.8\textwidth}{!}{
\includegraphics[width=6cm]{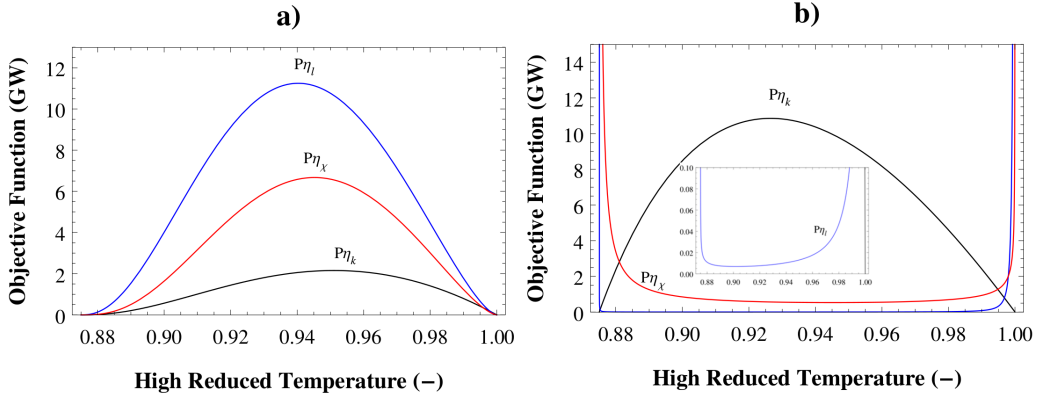}}
\caption{Graph of generalized commitment functions:
$k$\textendash Efficient Power, $l$\textendash Efficient
Power and $\chi$\textendash Efficient Power. By using the values, $\alpha=1\left[\textrm{MW}/\textrm{K}\right],\ \gamma=3,\ T_{h}=500\left[\textrm{K}\right]$, with a) $\ k=\ l=\chi=1.5$ and b) $\ k=\ -0.01, l=\ -2, \chi=-0.5$. Note in a) all the maximums are in different places, in spite of the generalization parameters are equal, they do not correspond at the same value of $a_{h}$. In b) all the parameters have different values because the domain of these functions is not the same.}
\label{fig:Generalizaciones}
\end{figure*}

\subsection{Energetics under the $k$--Efficient Power operation mode}
\label{sec:2.1}
By substituting the Eqs. \ref{eq:PeMEndo} and \ref{eq:EfieMEndo}
in the Eq. \ref{eq:kPotEfi}, we get the explicit expression for
the $k$--Efficient Power,

\begin{equation}
P\eta_{k}=T_{h}\alpha\left[1-a_{h}\right]\left[1-\frac{\tau}{a_{h}-\left(1-a_{h}\right)\gamma}\right]^{1+k}.\label{eq:kPotEfiEndo}
\end{equation}

Where $-1\leq k$, so that Eq. \ref{eq:kPotEfiEndo} has physical validity
\cite{LevarioArias}. From the Fig. \ref{fig:Generalizaciones}a, it
is possible to remark there exists an $a_{h}$ that maximizes the
$k$--Efficient Power, which is:

\begin{equation}
a_{h}^{mP\eta_{k}}\left[\gamma,\tau,k\right]=\frac{2\gamma-k\tau+\sqrt{\tau\left(4+4k+k^{2}\tau\right)}}{2\left(1+\gamma\right)}.\label{eq:ahMaxkPotEfiEndo}
\end{equation}

The process variables evaluated in this regime are:

\begin{equation}
P\left(\alpha,\gamma,\tau,T_{h},k\right)=T_{h}\alpha\frac{\left[\left(2+k\right)\sqrt{\tau}-\sqrt{4+k\left(4+k\tau\right)}\right]\left[2+k\tau-\sqrt{\tau}\sqrt{4+k\left(4+k\tau\right)}\right]}{2\left(1+\gamma\right)\left[k\sqrt{\tau}-\sqrt{4+k\left(4+k\tau\right)}\right]},\label{eq:PotMaxkPotEfi}
\end{equation}

\begin{equation}
\eta\left(\tau,k\right)=\frac{2+k\left(2-\tau\right)-\sqrt{\tau}\sqrt{4+k\left(4+k\tau\right)}}{2\left(1+\gamma\right)},\label{eq:EfiMaxkPotEfi}
\end{equation}
and
\begin{equation}
\Phi\left(\alpha,\gamma,\tau,T_{h},k\right)=T_{h}\alpha\frac{\sqrt{\tau}\left[2+k\tau-\sqrt{\tau}\sqrt{4+k\left(4+k\tau\right)}\right]^{2}}{2\left(1+\gamma\right)\left[\sqrt{4+k\left(4+k\tau\right)}-k\sqrt{\tau}\right]}.\label{eq:DisMaxkPotEfi}
\end{equation}

\begin{figure*}
\centering
\resizebox{0.85\textwidth}{!}{
\includegraphics[width=6cm]{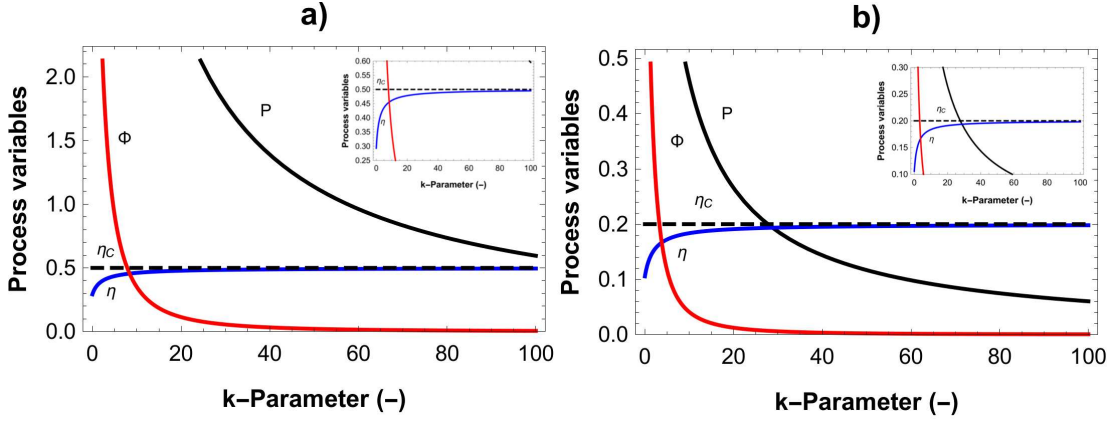}}
\caption{Energetics in the regime of maximum
$k$--Efficient Power, in which we have used values of: $\gamma=3,\ T_{h}=500\textrm{K},\ \alpha=1\left[\textrm{MW}/\textrm{K}\right].$
In Figure (a) we took $\tau=0.5$ and for (b) $\tau=0.8$. Where $\eta$ is the efficiency, $P$ is the power output in $\left[\textrm{GW}\right]$, 
$\Phi$ is the dissipation in $\left[\textrm{GW}\right]$ and $\eta_{C}$ the Carnot efficiency as the reversible limit.}
\label{fig:EnegereticaMaxkPotEfi}
\end{figure*}

From the Eq. \ref{eq:ahMaxkPotEfiEndo}, we note that this high reduced
temperature not only maximizes the $k$--Efficient Power, but it depends
on the generalization parameter. In Fig. \ref{fig:EnegereticaMaxkPotEfi} it
can be shown how the process variables are affected by the generalization
parameter $k$. However, the greater effect is reflected in the dissipation,
since increasing the value of $k$, it decreases faster than the power
output. On the other hand, the effect of this parameter makes the
efficiency reach a constant value, that is, the heat engines that
operate at maximum $k$--Efficient Power tend to the maximum efficiency
regime and therefore, the operation modes close to it are difficult
to achieve. We can also find the following limits for the process
variables in the maximum $k$--Efficient Power regime,

\begin{eqnarray}
\lim_{k\rightarrow\infty}P & = & 0,\nonumber \\
\lim_{k\rightarrow\infty}\eta & = & 1-\tau,\label{eq:LimkTInfEndo}\\
\lim_{k\rightarrow\infty}\Phi & = & 0.\nonumber 
\end{eqnarray}
 
\subsection{Energetics under the $l$--Efficient Power operation mode}
\label{sec:2.2}
\begin{figure*}
\centering
\resizebox{0.85\textwidth}{!}{
\includegraphics[width=6cm]{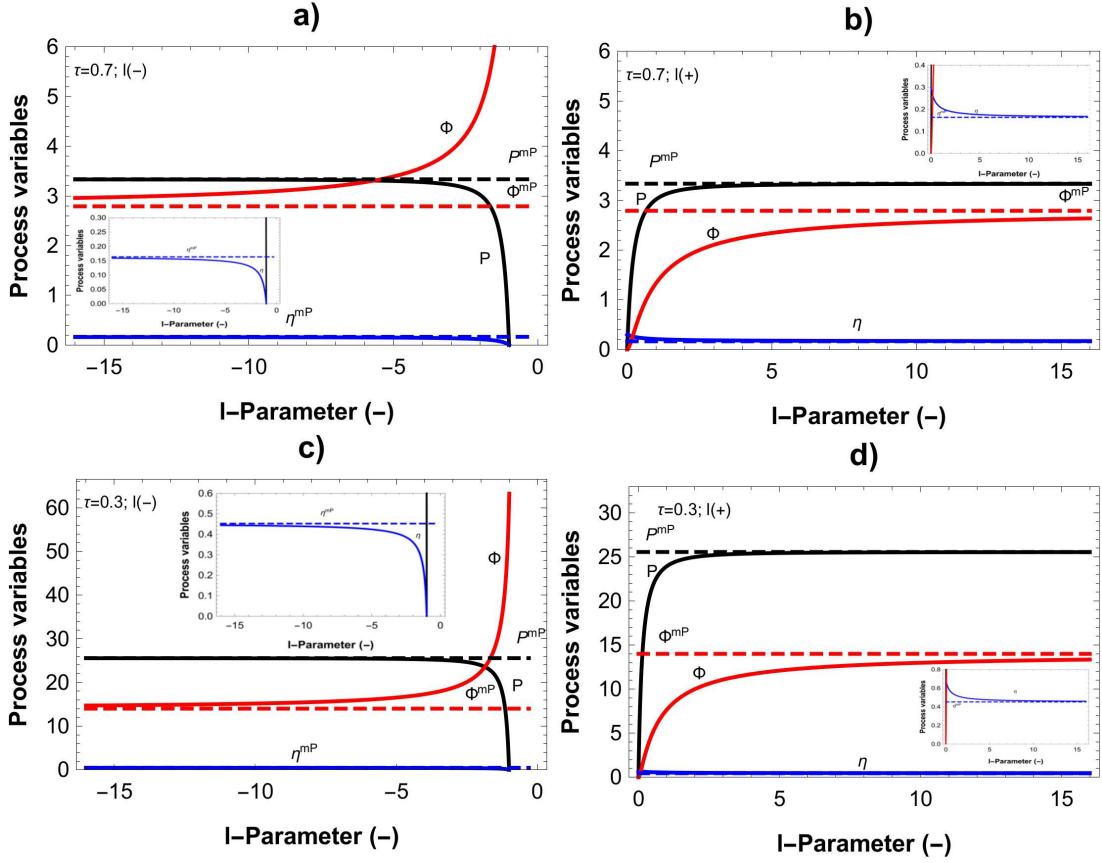}}
\caption{Energetics at the regime of maximum
$l$--Efficient Power for two values of $\tau$, in which we have used values of: 
$\gamma=3,\ T_{h}=500\textrm{K},\ \alpha=1\left[\textrm{MW}/\textrm{K}\right]$. In graphs a, b, c and d we show the effect of $a_{h1}$ and $a_{h2}$. With $P$ the power output in $\left[\textrm{GW}\right]$, 
$\Phi$ the dissipation in $\left[\textrm{GW}\right]$, $\eta$ the efficiency and the superscript $mP$ indicates the energetic of the system evaluated in 
the maximum power regime as the limit for each function.}
\label{fig:EnergeticaMaxlPotEfi}
\end{figure*}

The $l$--Efficient Power can be written as:

\begin{equation}
P\eta_{l}=\left[T_{h}\alpha\left(1-a_{h}\right)\right]^{l}\left[1-\frac{\tau}{a_{h}+\gamma\left(a_{h}-1\right)}\right]^{l+1},\label{eq:lPotEfiEndo}
\end{equation}
where $l\in\left(-\infty,-1\right]\cup\left(0,\infty\right)$ otherwise
it would not exhibit attainable physical properties. In the same way,
this commitment function has two different high reduced temperatures
that maximize it (Fig. \ref{fig:EnergeticaMaxlPotEfi}), the first
one:
\begin{equation}
a_{h1}^{mP\eta_{l}}\left[\gamma,\tau,l\right]=\frac{2l\gamma-\tau+\sqrt{\tau\left[4l\left(1+l\right)+\tau\right]}}{2l\left(1+\gamma\right)},\label{eq:ah1MaxlPotEfi}
\end{equation}
represents a maximum located in the zone where the energy converter
operates as a heat engine for $l\in\left(0,\infty\right)$. Beneath
these same conditions, the second high reduced temperature
\begin{equation}
a_{h2}^{mP\eta_{l}}\left[\gamma,\tau,l\right]=\frac{2l\gamma-\tau-\sqrt{\tau\left[4l\left(1+l\right)+\tau\right]}}{2l\left(1+\gamma\right)}\label{eq:ah2MaxlPotEfi}
\end{equation}
depicts a minimum (see Fig. \ref{fig:Generalizaciones}b), which is in the region where the energy converter does not operates as a heat engine. With this operation regimen, the process variables
become,

\[
P_{1}\left(\alpha,\gamma,\tau,T_{h},l\right)=T_{h}\alpha\frac{\left[\left(1+2l\right)\sqrt{\tau}+\sqrt{4l\left(1+l\right)+\tau}\right]\left[2l+\tau+\sqrt{\tau}\sqrt{4l\left(1+l\right)+\tau}\right]}{2l\left(1+\gamma\right)\left[\sqrt{\tau}+\sqrt{4l\left(1+l\right)+\tau}\right]},
\]

\[
\eta_{1}\left(\tau,l\right)=\frac{2\left(1+l\right)-\tau+\sqrt{\tau}\sqrt{4l\left(1+l\right)+\tau}}{2\left(1+l\right)}
\]
and
\[
\Phi_{1}\left(\alpha,\gamma,\tau,T_{h},l\right)=-T_{h}\alpha\frac{\sqrt{\tau}\left[2l+\tau+\sqrt{\tau}\sqrt{4l\left(1+l\right)+\tau}\right]^{2}}{2l\left(1+\gamma\right)\left[\sqrt{\tau}+\sqrt{4l\left(1+l\right)+\tau}\right]}.
\]

While

\[
P_{2}\left(\alpha,\gamma,\tau,T_{h},l\right)=T_{h}\alpha\frac{\left[\left(1+2l\right)\sqrt{\tau}-\sqrt{4l\left(1+l\right)+\tau}\right]\left[2l+\tau-\sqrt{\tau}\sqrt{4l\left(1+l\right)+\tau}\right]}{2l\left(1+\gamma\right)\left[\sqrt{\tau}-\sqrt{4l\left(1+l\right)+\tau}\right]},
\]

\[
\eta_{2}\left(\tau,l\right)=\frac{2\left(1+l\right)-\tau-\sqrt{\tau}\sqrt{4l\left(1+l\right)+\tau}}{2\left(1+l\right)}
\]
and
\[
\Phi_{2}\left(\alpha,\gamma,\tau,T_{h},l\right)=T_{h}\alpha\frac{\sqrt{\tau}\left[2l+\tau-\sqrt{\tau}\sqrt{4l\left(1+l\right)+\tau}\right]^{2}}{2l\left(1+\gamma\right)\left[\sqrt{4l\left(1+l\right)+\tau}-\sqrt{\tau}\right]}.
\]

When $l\in\left(-\infty,-1\right]$, the Eq. \ref{eq:ah1MaxlPotEfi},
still represents a maximum, though. It is located in the zone where
the thermal machine does not work as a heat engine. Analogously, it
happens with the Eq. \ref{eq:ah2MaxlPotEfi}. 

In Fig. \ref{fig:EnergeticaMaxlPotEfi} we can see two regions in
which the process variables operationally describe the energy converter
as a heat engine. These regions are separated by the point that characterizes
the maximum power output regime. The behavior
of process variables tend to very specific limits, given by:

\begin{eqnarray}
\lim_{l\rightarrow\pm\infty}P & = & T_{h}\alpha\frac{\left(\sqrt{\tau}-1\right)^{2}}{1+\gamma}\nonumber \\
\lim_{l\rightarrow\pm\infty}\eta & =1- & \sqrt{\tau}\label{eq:LimlTInf}\\
\lim_{l\rightarrow\pm\infty}\Phi & = & T_{h}\alpha\frac{\sqrt{\tau}\left(\sqrt{\tau}-1\right)^{2}}{1+\gamma}.\nonumber 
\end{eqnarray}

This function has a clear disadvantage with respect to the $k$--Efficient
Power, if you want to study the real energy converters (heat engines),
because this function does not include a point which is physically
accessible: the maximum power output regime.

\subsection{Energetics under the $\chi$-Efficient Power operation mode}
\label{sec:2.3}
In the case of $\chi$--Efficient Power. The model of the function
has the following shape,

\begin{equation}
P\eta_{\chi}=\left[\frac{T_{h}\alpha\left(1-a_{h}\right)\left\{ \gamma+\tau-a_{h}\left(1+\gamma\right)\right\} ^{2}}{\left(a_{h}+\gamma\left(a_{h}-1\right)\right)^{2}}\right]^{\chi}.\label{eq:mPotEfiEndo}
\end{equation}

This function also presents an extreme value at the high reduced temperature
and that maximizes it (see Fig. \ref{fig:Generalizaciones}a). This value is:

\begin{equation}
a_{h}^{mP\eta_{\chi}}[\gamma,\tau]=\frac{2\gamma-\tau+\sqrt{\tau\left(8+\tau\right)}}{2\left(1+\gamma\right)},\label{eq:ahMaxmpotEfi}
\end{equation}

As can be beheld, in Eq. \ref{eq:ahMaxmpotEfi}, this high reduced
temperature that characterizes the optimal regime for this generalization,
does not depend on the parameter $\chi$. It coincides with the same
results that has already been obtained for Efficient Power \cite{Yilmaz}.

\subsection{A comparison between the different generalization proposals}
\label{sec:2.4}
The initial analysis presented below has the objective of comparing
the mathematical forms of the generalization proposals, whose generalization
parameter appears explicitly in the energetics of every heat engine,
in order to show some advantages of each one of their operation modes.
\begin{figure*}
\centering
\resizebox{0.85\textwidth}{!}{
\includegraphics[width=6cm]{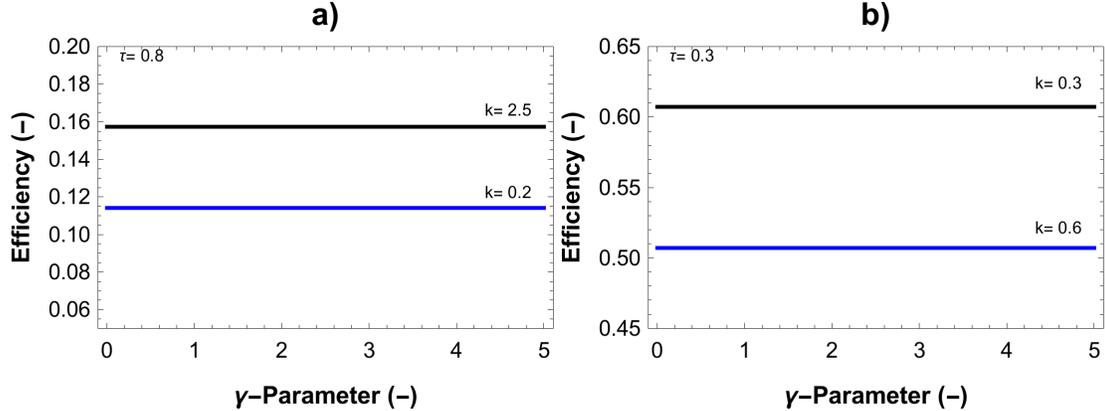}}
\caption{Comparison between the efficiencies
at maximum $k$--Efficient Power for four different values of $k$ and two values of $\tau$.}
\label{fig:Comparativa-entra-las}
\end{figure*}
It is possible to detect
certain differences in relation to the optimal regimes that each one
of them let us to reach. In Fig. \ref{fig:Comparativa-entra-las}
the efficiencies are completely different for the values that can
be assumed by $l$ and $k$ respectively.

In general, the generalization parameters play a big role in the optimization
processes for energy converters. For instance, from the graphs of
Fig. \ref{fig:Comparativa-entra-las} we can observe the behavior
of efficiency for two different values of $\tau$. In particular,
the graphs a) and b) of Fig. \ref{fig:Comparativa-entra-las} make
evident the experimental fact that occurs in the mesoscopic energy
converters, which operate within very small temperature gradients $(T_{c}\thickapprox T_{h})$
and therefore, their efficiencies are low \cite{ZhouChenDingSun16,HumphreyNewburyTaylorLinke02,NakpathomkunXuLinke10}.
This is because the system become saturated and the energy conversion
processes stop. In contrast with macroscopic systems, where the restriction
in the temperature relationship does not exist, since the energy reservoirs
are considered large enough as for the thermal machines to operate
continuously and achieve greater efficiencies. Another consequence
of implementing the parameters of generalization in objective functions
is the possibility of improving efficiency and in general any of the
process variables, \textit{i.e.}, they enable to modify the heat engine performance.

\section{Effects on the construction and improvements in some power plants}
\label{sec:3}
The operation of a thermal machine takes into account a significant
number of variables, since most of them are related not only to the
design of the energy converter but also to the mode of operation that
can be achieved. From the CA model we denote the parameter $\gamma$
as the relationship exists between the thermal conductances, which
modulate the energy exchange with the reservoirs. While the relationship
between the temperatures of the reservoirs $\tau$ is related to the
operating range of this converter, \textit{i.e.}, it bounds the capacity of
the system to convert energy. Due to the above, we will consider these
two parameters as control elements in the performance of a heat engine, they modulate the amount of energy that gets in and gets
out the system. On the other hand, as we have already pointed out
the parameter $k$ allows to describe the gains or losses of energy
which are reflected in the process variables of a heat engine.

In \cite{LevarioArias} has been shown the generalization parameter
$k$ can help us to reach any operation mode that allows an energy converter
to work as a heat engine. This has certain advantages, because the $k$
generalization parameter indicates whether the thermal machine operates
in a low efficiency (LE) and high dissipation (HD) zone or in one
of high efficiency (HE) and low dissipation (LD). We consider those operation modes whose efficiency values are bounded between the efficiency in the maximum power regime and Carnot regime, this operation modes belong to HE zone and therefore their dissipation values are in the LD zone.

If we have any of the process variables (Eqs. \ref{eq:PeMEndo}, \ref{eq:EfieMEndo} and
\ref{eq:DisEMEndo}), we can establish, through the $k$--Efficient Power regime, the relations that exist
between the parameter $k$ and the energetic of the heat engine, given by:

\begin{eqnarray}
k_{LE}\left(\alpha,\gamma,T_{h},\tau,P\right) & = & \frac{P^{2}\left(\gamma+1\right)^{2}+T_{h}\alpha\left(1-\tau\right)\left[T_{h}\alpha\left(1-\tau\right)-\varLambda\right]-P\left(1+\gamma\right)\left[2T_{h}\alpha\left(1+\tau\right)+\varLambda\right]}{2PT_{h}\alpha\tau\left(1+\gamma\right)}\label{eq:kenfundepot}\\
k_{HE}\left(\alpha,\gamma,T_{h},\tau,P\right) & = & \frac{P^{2}\left(\gamma+1\right)^{2}+T_{h}\alpha\left(1-\tau\right)\left[T_{h}\alpha\left(1-\tau\right)+\varLambda\right]-P\left(1+\gamma\right)\left[2T_{h}\alpha\left(1+\tau\right)+\varLambda\right]}{2PT_{h}\alpha\tau\left(1+\gamma\right)}\label{eq:kenfundepothe}\\
k\left(\tau,\eta\right) & = & \frac{1-\tau-\eta\left(2-\eta\right)}{\left(1-\eta\right)\left(\eta+\tau-1\right)}\label{eq:kenfundeefic}\\
k_{HE}\left(\alpha,\gamma,T_{h},\tau,\Phi\right) & = & \frac{\left(1+\gamma\right)\Phi\left[\Xi-\left(1+\gamma\right)\sqrt{\Phi}\right]+T_{h}\alpha\tau\left[\left(1-\tau\right)\Xi-\left(1+\gamma\right)\left(3+\tau\right)\sqrt{\Phi}\right]}{2T_{h}\alpha\left(1+\gamma\right)\sqrt{\Phi}\tau^{2}}\label{eq:kenfundedis}\\
k_{LE}\left(\alpha,\gamma,T_{h},\tau,\Phi\right) & = & -\frac{\left(1+\gamma\right)\Phi\left[\Xi+\left(1+\gamma\right)\sqrt{\Phi}\right]+T_{h}\alpha\tau\left[\left(\tau-1\right)\Xi+\left(1+\gamma\right)\left(3+\tau\right)\sqrt{\Phi}\right]}{2T_{h}\alpha\left(1+\gamma\right)\sqrt{\Phi}\tau^{2}}\label{eq:kenfundedisle}
\end{eqnarray}
where

\[
\begin{array}{c}
\Lambda=\sqrt{P^{2}\left(1+\gamma\right)^{2}+T_{h}^{2}\alpha^{2}\left(1-\tau\right)^{2}-2PT_{h}\alpha\left(1+\gamma\right)\left(1+\tau\right)}\\
\Xi=\sqrt{\left(1+\gamma\right)\left(4T_{h}\alpha\tau+\Phi\left[1+\gamma\right]\right)}
\end{array}.
\]

The remarkable of the Eqs. \ref{eq:kenfundepot} and \ref{eq:kenfundeefic}
is manifested through an energy analysis applied to a set of 6 power  plants (2 of monocycle, 2 of combined cycle and 2 nuclear).
In Table \ref{tab:Valorelk} the reported manufactured values of: reservoir temperatures
($T_{h}$ and $T_{c}$), efficiency ($\eta$) and operating power
output ($P$) for each of them are presented. With this data \cite{Plantas01,Plantas02}, we use
the Eq. \ref{eq:kenfundeefic} to know the value of $k$ that characterizes
the operation mode of each power plant, assuming a fixed relationship
between the conductances. In addition, with the operating
power output and Eq. \ref{eq:kenfundepot}, the value of thermal conductance
$\alpha$ is estimated. Finally, from Eq. \ref{eq:ahMaxkPotEfiEndo}, the high reduced temperature value $(a_{h})$ that characterizes
the operation of each plant is reckoned up, with all these
parameters, we are able to estimate the dissipated energy of each power 
plant. In the same Table \ref{tab:Valorelk}, it is shown a comparison
of the values that these plants should have if they operate in any
of the already well-known regimes (maximum power output $\left(MP\right)$,
maximum efficiency $\left(M\eta\right)$, maximum ecological function
$\left(ME\right)$, maximum efficient power $\left(MP_{\eta}\right)$
and maximum generalized ecological function $\left(ME_{g}\right)$).

\begin{table}
\begin{centering}
\caption{Operation data for different commercial power plants \cite{Plantas01,Plantas02}. A comparison of these reported power output
($P_{r}$) and efficiency ($\eta_{r}$) against the values of these
process variables would reach for different operation modes already
known: $MP$, $M\eta$, $MP\eta$, $ME$, y $ME_{g}$. The worth of
$k$ that allows to characterize the operation regime of the different
power plants, are associated with their corresponding
high reduced temperature $(a_{h})$, as well as the dissipation that
is generated for their respective operating regime. An arbitrary conductances
ratio has been assumed $\left(\gamma=3\right)$ and in each case the
value of the conductance $\alpha$ has been calculated which allows
to obtain the value of the reported power output.}
\label{tab:Valorelk}
{\scriptsize{}}%
\begin{tabular}{|c|c|c|c|c|c|c|c|c|c|c|c|c|}
\hline 
{\scriptsize{}Power Plant} & \multicolumn{2}{c|}{{\scriptsize{}Almaraz II (A)}} & \multicolumn{2}{c|}{{\scriptsize{}Cofrentes (C)}} & \multicolumn{2}{c|}{{\scriptsize{}West Thurrock (WT)}} & \multicolumn{2}{c|}{{\scriptsize{}Larderello (L)}} & \multicolumn{2}{c|}{{\scriptsize{}Toshiba (T)}} & \multicolumn{2}{c|}{{\scriptsize{}Alstom (Al)}}\tabularnewline
 & \multicolumn{2}{c|}{{\scriptsize{}(PWR, Spain, 83)}} & \multicolumn{2}{c|}{{\scriptsize{}(BWR, Spain, 84)}} & \multicolumn{2}{c|}{{\scriptsize{}(Uk, 62)}} & \multicolumn{2}{c|}{{\scriptsize{} (Italy,64)}} & \multicolumn{2}{c|}{{\scriptsize{}(109FA, 04)}} & \multicolumn{2}{c|}{{\scriptsize{}(ka26-1)}}\tabularnewline
\hline 
{\scriptsize{}Reported Data} & {\scriptsize{}$T_{h}[K]$} & {\scriptsize{}$T_{c}[K]$} & {\scriptsize{}$T_{h}[K]$} & {\scriptsize{}$T_{c}[K]$} & {\scriptsize{}$T_{h}[K]$} & {\scriptsize{}$T_{c}[K]$} & {\scriptsize{}$T_{h}[K]$} & {\scriptsize{}$T_{c}[K]$} & {\scriptsize{}$T_{h}[K]$} & {\scriptsize{}$T_{c}[K]$} & {\scriptsize{}$T_{h}[K]$} & {\scriptsize{}$T_{c}[K]$}\tabularnewline
\cline{2-13} 
 & {\scriptsize{}600} & {\scriptsize{}290} & {\scriptsize{}562} & {\scriptsize{}289} & {\scriptsize{}838} & {\scriptsize{}298} & {\scriptsize{}523} & {\scriptsize{}353} & {\scriptsize{}1573} & {\scriptsize{}303} & {\scriptsize{}1398} & {\scriptsize{}288}\tabularnewline
\cline{2-13} 
 & {\scriptsize{}$P_{r}\left[GW\right]$} & {\scriptsize{}$\eta_{r}$} & {\scriptsize{}$P_{r}\left[GW\right]$} & {\scriptsize{}$\eta_{r}$} & {\scriptsize{}$P_{r}\left[GW\right]$} & {\scriptsize{}$\eta_{r}$} & {\scriptsize{}$P_{r}\left[GW\right]$} & {\scriptsize{}$\eta_{r}$} & {\scriptsize{}$P_{r}\left[GW\right]$} & {\scriptsize{}$\eta_{r}$} & {\scriptsize{}$P_{r}\left[GW\right]$} & {\scriptsize{}$\eta_{r}$}\tabularnewline
\cline{2-13} 
 & {\scriptsize{}1.044} & {\scriptsize{}0.35} & {\scriptsize{}1.092} & {\scriptsize{}0.34} & {\scriptsize{}1.240} & {\scriptsize{}0.36} & {\scriptsize{}0.150} & {\scriptsize{}0.16} & {\scriptsize{}0.342} & {\scriptsize{}0.48} & {\scriptsize{}0.410} & {\scriptsize{}0.57}\tabularnewline
\hline 
{\scriptsize{}$P^{MP}; \eta^{MP}$} & {\scriptsize{}1.0806} & {\scriptsize{}0.30} & {\scriptsize{}1.1638} & {\scriptsize{}0.28} & {\scriptsize{}1.2631} & {\scriptsize{}0.4} & {\scriptsize{}0.1519} & {\scriptsize{}0.18} & {\scriptsize{}0.3563} & {\scriptsize{}0.56} & {\scriptsize{}0.4118} & {\scriptsize{}0.55}\tabularnewline
\hline 
{\scriptsize{}$P^{MP\eta}; \eta^{MP\eta}$} & {\scriptsize{}0.9944} & {\scriptsize{}0.37} & {\scriptsize{}1.0683} & {\scriptsize{}0.35} & {\scriptsize{}1.1759} & {\scriptsize{}0.48} & {\scriptsize{}0.1378} & {\scriptsize{}0.23} & {\scriptsize{}0.3379} & {\scriptsize{}0.64} & {\scriptsize{}0.3899} & {\scriptsize{}0.62}\tabularnewline
\hline 
{\scriptsize{}$P^{ME}; \eta^{ME}$} & {\scriptsize{}0.8997} & {\scriptsize{}0.40} & {\scriptsize{}0.9617} & {\scriptsize{}0.38} & {\scriptsize{}1.0878} & {\scriptsize{}0.51} & {\scriptsize{}0.1211} & {\scriptsize{}0.25} & {\scriptsize{}0.3229} & {\scriptsize{}0.66} & {\scriptsize{}0.3715} & {\scriptsize{}0.65}\tabularnewline
\hline 
{\scriptsize{}$P^{ME_{g}}; \eta^{ME_{g}}$} & {\scriptsize{}0.8127} & {\scriptsize{}0.42} & {\scriptsize{}0.8749} & {\scriptsize{}0.39} & {\scriptsize{}0.9525} & {\scriptsize{}0.54} & {\scriptsize{}0.1140} & {\scriptsize{}0.26} & {\scriptsize{}0.2709} & {\scriptsize{}0.71} & {\scriptsize{}0.3128} & {\scriptsize{}0.69}\tabularnewline
\hline 
{\scriptsize{}$\alpha\left[MW/K\right]$} & \multicolumn{2}{c|}{{\scriptsize{}77.5543}} & \multicolumn{2}{c|}{{\scriptsize{}103.5040}} & \multicolumn{2}{c|}{{\scriptsize{}36.9997}} & \multicolumn{2}{c|}{{\scriptsize{}36.4921}} & \multicolumn{2}{c|}{{\scriptsize{}2.8779}} & \multicolumn{2}{c|}{{\scriptsize{}3.9509}}\tabularnewline
\hline 
{\scriptsize{}Generalization Data} & \multicolumn{2}{c|}{{\scriptsize{}$k$}} & \multicolumn{2}{c|}{{\scriptsize{}$k$}} & \multicolumn{2}{c|}{{\scriptsize{}$k$}} & \multicolumn{2}{c|}{{\scriptsize{}$k$}} & \multicolumn{2}{c|}{{\scriptsize{}$k$}} & \multicolumn{2}{c|}{{\scriptsize{}$k$}}\tabularnewline
\cline{2-13} 
 & \multicolumn{2}{c|}{{\scriptsize{}0.5615}} & \multicolumn{2}{c|}{{\scriptsize{}0.8174}} & \multicolumn{2}{c|}{{\scriptsize{}-0.2966}} & \multicolumn{2}{c|}{{\scriptsize{}-0.2211}} & \multicolumn{2}{c|}{{\scriptsize{}-0.4569}} & \multicolumn{2}{c|}{{\scriptsize{}0.2192}}\tabularnewline
\cline{2-13} 
 & \multicolumn{2}{c|}{{\scriptsize{}$a_{h}$}} & \multicolumn{2}{c|}{{\scriptsize{}$a_{h}$}} & \multicolumn{2}{c|}{{\scriptsize{}$a_{h}$}} & \multicolumn{2}{c|}{{\scriptsize{}$a_{h}$}} & \multicolumn{2}{c|}{{\scriptsize{}$a_{h}$}} & \multicolumn{2}{c|}{{\scriptsize{}$a_{h}$}}\tabularnewline
\cline{2-13} 
 & \multicolumn{2}{c|}{{\scriptsize{}0.9359}} & \multicolumn{2}{c|}{{\scriptsize{}0.9448}} & \multicolumn{2}{c|}{{\scriptsize{}0.8889}} & \multicolumn{2}{c|}{{\scriptsize{}0.9509}} & \multicolumn{2}{c|}{{\scriptsize{}0.8426}} & \multicolumn{2}{c|}{{\scriptsize{}0.8698}}\tabularnewline
\hline 
{\scriptsize{}$\Phi\left[GW\right]$} & \multicolumn{2}{c|}{{\scriptsize{}0.4971}} & \multicolumn{2}{c|}{{\scriptsize{}0.4682}} & \multicolumn{2}{c|}{{\scriptsize{}0.9796}} & \multicolumn{2}{c|}{{\scriptsize{}0.1547}} & \multicolumn{2}{c|}{{\scriptsize{}0.2333}} & \multicolumn{2}{c|}{{\scriptsize{}0.1611}}\tabularnewline
\hline 
\end{tabular}
\par\end{centering}{\scriptsize \par}

\end{table}

\begin{figure*}
\centering
\resizebox{1.0\textwidth}{!}{
\includegraphics[width=6cm]{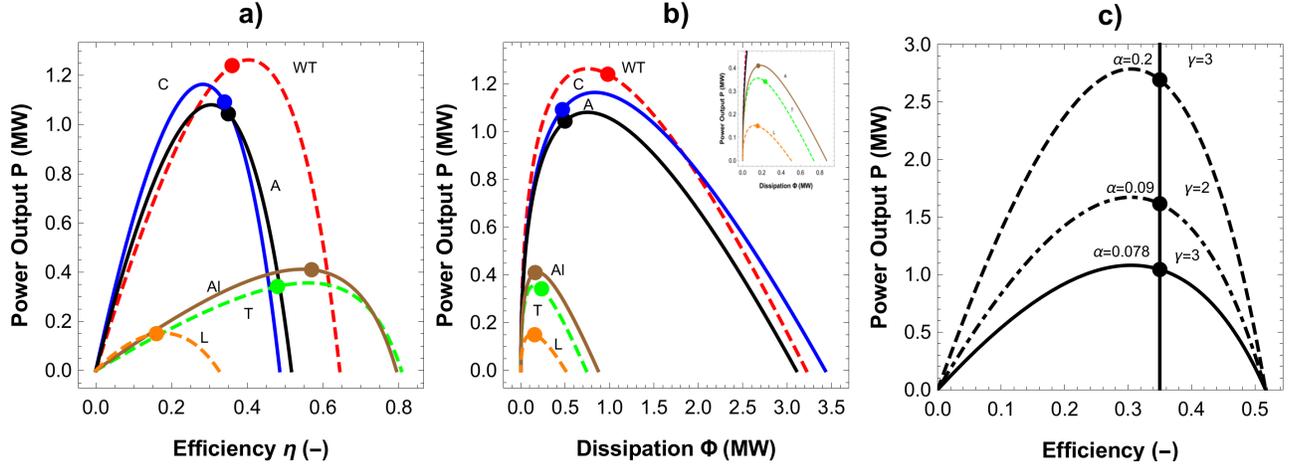}}
\caption{a) Curves of power output \textit{vs.} efficiency
and b) curves of power output \textit{vs.} dissipation for each of the power
plants reported in Table \ref{tab:Valorelk}. In solid lines the plants
that operate in the HE zone are shown [Almaraz II (black, A), Cofrentes (blue, C) and Alston (brown, Al)], while the dashed
lines show those ones that work in the LE zone [West Thurrock (red, WT), Toshiba (green, T) and Larderello (orange, L)]. In c)
the energetic behavior that Almaraz II achieves for different values
of the control parameters $\alpha$ and $\gamma$ is exhibited.}
\label{fig:CurvasPotEfiPlantas}
\end{figure*}

From Eqs. \ref{eq:PotMaxkPotEfi} and \ref{eq:EfiMaxkPotEfi}, we can use the control parameters to tune the performance of a heat engine. On the other hand, $\tau$ can be
associated with a kind of thermal potential, because it accounts for
the length of the temperature difference between the external energy
reservoirs. As we can note, from Fig. \ref{fig:CurvasPotEfiPlantas} and Table \ref{tab:Valorelk}, 
$\tau$ is directly related to the efficiencies that this set of power plants can perform, such is the case of Larderello (L), whose temperature difference between the reservoirs is small, this minimizes the gradient responsible for promoting the energy flux through the converter, \textit{i.e.}, a small number of cycles is generated in the heat engine compared to Cofrentes (C) plant, whose value of $\tau$ is the closest to Larderello one, so that Larderello's power output is low. Due to the energy conservation, a large amount of heat flux must be transferred to the reservoir at $T_{c}$. From the Eqs. \ref{eq:PeMEndo}, \ref{eq:EfieMEndo} and \ref{eq:DisEMEndo}, it is had that for those operation modes whose efficiency is below a half of the Carnot efficiency, the system will dissipate more energy than the power output obtained.
 
In the case of the conductances ($\alpha$,$\beta$) and their ratio $\left(\gamma\right)$,
their greatest effect is reflected in the power output that this kind
of converter can deliver as useful energy. According to the CA model, all of power plants express a relationship between
the reported power and the value of the parameters
($\gamma$,$\alpha$) (see Table \ref{tab:Valorelk}). However, when modifying any of these two parameters,
the power output can increase or decrease, depending on the value
that they acquire (see Fig. \ref{fig:CurvasPotEfiPlantas}).

This indicates that the reported values of power output and efficiency of
a power plant are not enough information to think of modifying
the operation of it. Such is the case of some combined cycle plants,
Toshiba 109FA of 2004 (T) it has an efficiency 
similar to Alston ka26-1 (Al) but with less power output. Although the first one reports a relatively high
efficiency compared to the other one, it is operating in the LE
and HD zone, while the second one operates in the HE and LD zone (Fig.
\ref{fig:CurvasPotEfiPlantas}a). In general, this behavior is an evidence of the operating mechanisms
to which the power plants are undergone, \textit{i.e.}, the operation of
each one defines, from the energy point of view, a family of curves
in the configuration space ($\alpha$,$\gamma$, $\tau$, $P$ and
$\eta$), it fixes to a well-defined point that represents
a characteristic operation regime. From the data for the plants, we can select a specific value of $\tau$,
which reduces the number of configurations for each one. It is verified,
for example in Almaraz II, that it is possible to assume $\alpha$ or $\gamma$, and then adjust either of the two, so that the reported power output will be reached (Fig. \ref{fig:CurvasPotEfiPlantas}b).
Although the family of curves that allow us to attain the reported power
output is still large, for the assumed parameter ($\gamma$),
we are able to obtain only one that adjusts to the value of such power output . Thus, the new configuration is completed and the result is a curve with any accessible operation
mode.

\begin{figure*}
\centering
\resizebox{1.0\textwidth}{!}{
\includegraphics[width=6cm]{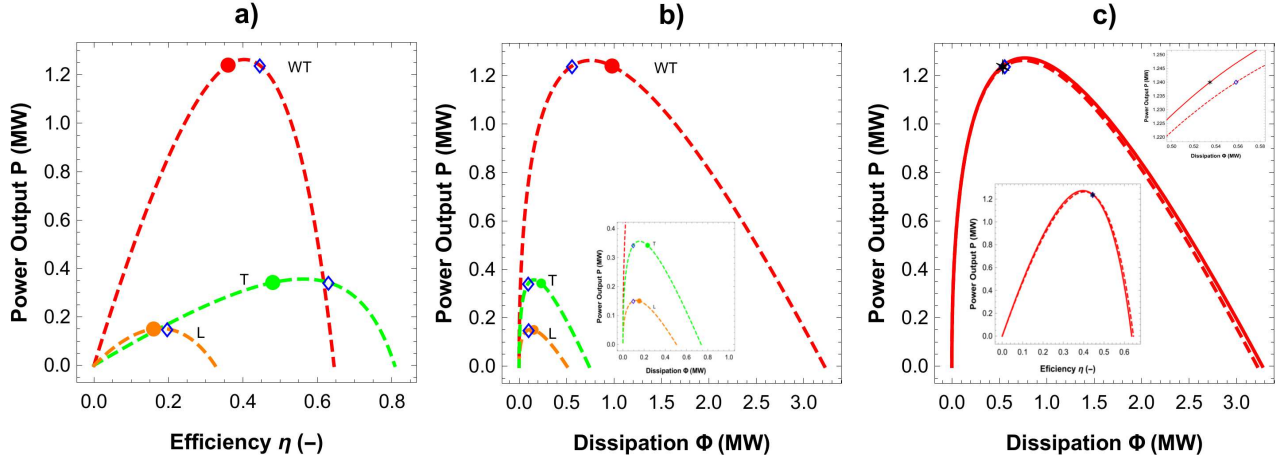}}
\caption{a) Power output \textit{vs.} efficiency curves
for the plants [West Thurrock (red, WT), Toshiba (green, T) and Larderello (orange,L)] whose operating modes are in the LE zone (marked with points), the new operation mode 
to which each plant can access is also shown, with the same power output but in the HE zone (marked with diamonds). 
While in b) Power output \textit{vs.} dissipation curves for the same plants are shown.  In c) the power output \textit{vs.} 
dissipation and power output \textit{vs.} efficiency curves are shown for the original WT plant configuration 
(dashed line), as well as its corresponding curves to the restructuring
configuration (solid line). In addition, the original operation point marked with a diamond, the best efficiency under the same configuration and
the restructuring condition are highlighted with a star.}
\label{fig:MejoraEnEnergetica}
\end{figure*}

In Fig. \ref{fig:CurvasPotEfiPlantas}a, it is observed
that some power plants already operate in the HE and LD zone. On the contrary, there are some of them which are in the HD
and LE zone, and it is possible to find the restructuring conditions that allow them to operate in an optimal configuration, producing the same power
output, a gain in efficiency and a decrease in dissipated energy.
To accomplish the above, it is important to find the physically achievable
efficiency values for a given power output. Then, we calculate $k$
through Eq. \ref{eq:kenfundepothe} and substitute
it in the expression for the efficiency (Eq. \ref{eq:EfiMaxkPotEfi}). And it is pointed out that the values of $k>0$ ($k^{\ast}$)
are associated with an efficiency located in the
HE and LD zone ($\eta^{\ast}$), while the values of $-1<k<0$ leads to an efficiency in the HD and LE zone (see
Fig. \ref{fig:MejoraEnEnergetica}a). On the other hand, $k$ is
related with the value of $a_{h}$ that characterizes
each operation mode. We noted,
if $a_{h}=a_{h}^{\ast}$ is within the HE and LD zone, is longer than the value acquired in the HD and LE zone
(Fig. \ref{fig:CurvasPotEfiPlantas}). Then,
the only way to establish the restructuring condition under the CA model,
at the same point in the configuration space, is modifying the reservoir
temperature $T_{h}$. This leads us to look for
a different configuration from the original one (curve) and that condition
guarantees the existence of the reported power output and improved
efficiency point, \textit{i.e.}, we assume that $T_{hw}$ remains constant and the relationship must be satisfied,

\begin{equation}
\frac{a_{h}}{a_{h}^{\ast}}=\frac{T_{h}^{\ast}}{T_{h}}.\label{eq:RTM}
\end{equation}
Note that $T_h^{\ast}<T_h$. By assuming that the temperature $T_{c}$ keeps
unchanged, we can now find a new $\tau=\tau^{RC}$.
Analogously, to calculate some data from the Table \ref{tab:Valorelk},
we take $\tau^{RC}$, the reported power output
$P_{r}$ and the efficiency $\eta^{\ast}$ in
order to derive new conditions on $\gamma$ and
$\alpha$, when we solve the equation
system:

\begin{eqnarray}
\eta^{\ast} & = & \eta\left(\gamma^{RC},\tau^{RC},a_{h}^{\ast}\right),\nonumber \\
P & = & P\left(\alpha^{RC},\gamma^{RC},\tau^{RC},T_{h}^{\ast},a_{h}^{\ast}\right).\label{eq:SEdRE}
\end{eqnarray}

Where the $RC$ index refers to the restructuring condition. The
parameters ($\alpha^{RC},\gamma^{RC},\tau^{RC},P$ and $\ensuremath{\eta^{\ast}}$)
represent a new configuration (restructuring configuration) in which
the operation mode of each plant is within the HE and LD zone. As
can be seen in Table \ref{tab:MejorasALaEficiencia}, the advantage
we find is less dissipation during the operation at the restructuring
condition. 

\begin{table}
\begin{centering}
\caption{A comparison between the original
operating modes, the same ones with the best efficiency under the
original configuration and the new operating regimes within the restructuring
condition for the WT, L and T power plants.}
\label{tab:MejorasALaEficiencia}
\begin{tabular}{|c|c|c|c|c|c|c|c|c|c|c|c|}
\hline 
{\tiny{}Plant} & {\tiny{}$k\notin HE$} & {\tiny{}$k\in HE$} & {\tiny{}$T_{h}/T_{h}^{\ast}$} & \multicolumn{2}{c|}{{\tiny{}RC}} & \multicolumn{2}{c|}{{\tiny{}$PV\notin HE$}} & \multicolumn{2}{c|}{{\tiny{}$PV\in HE$}} & \multicolumn{2}{c|}{{\tiny{}$PV^{RC}$}}\tabularnewline
\hline 
{\tiny{}West Thurrock (WT)} & {\tiny{}-0.2966} & {\tiny{}0.4217} & {\tiny{}0.9768} & {\tiny{}$\alpha^{RC}\left[MW/K\right]$} & {\tiny{}37.8776} & {\tiny{}$P\left[GW\right]$} & {\tiny{}1.24} & {\tiny{}$P\left[GW\right]$} & {\tiny{}1.24} & {\tiny{}$P\left[GW\right]$} & {\tiny{}1.24}\tabularnewline
\cline{5-12} 
 &  &  &  & {\tiny{}$^{\gamma^{RC}}$} & {\tiny{}2.8313} & {\tiny{}$\eta$} & {\tiny{}0.36} & {\tiny{}$\eta^{\ast}$} & {\tiny{}0.44} & {\tiny{}$\eta$} & {\tiny{}0.44}\tabularnewline
\cline{5-12} 
 &  &  &  & {\tiny{}$\tau^{RC}$} & {\tiny{}0.3640} & {\tiny{}$\Phi\left[GW\right]$} & {\tiny{}0.98} & {\tiny{}$\Phi^{\ast}\left[GW\right]$} & {\tiny{}0.558} & {\tiny{}$\Phi\left[GW\right]$} & {\tiny{}0.535}\tabularnewline
\hline 
{\tiny{}Larderello (L)} & {\tiny{}-0.2211} & {\tiny{}0.2838} & {\tiny{}0.9905} & {\tiny{}$\alpha^{RC}\left[MW/K\right]$} & {\tiny{}36.8422} & {\tiny{}$P\left[GW\right]$} & {\tiny{}0.15} & {\tiny{}$P\left[GW\right]$} & {\tiny{}0.15} & {\tiny{}$P\left[GW\right]$} & {\tiny{}0.15}\tabularnewline
\cline{5-12} 
 &  &  &  & {\tiny{}$^{\gamma^{RC}}$} & {\tiny{}2.7986} & {\tiny{}$\eta$} & {\tiny{}0.16} & {\tiny{}$\eta^{\ast}$} & {\tiny{}0.20} & {\tiny{}$\eta$} & {\tiny{}0.20}\tabularnewline
\cline{5-12} 
 &  &  &  & {\tiny{}$\tau^{RC}$} & {\tiny{}0.6814} & {\tiny{}$\Phi\left[GW\right]$} & {\tiny{}0.15} & {\tiny{}$\Phi^{\ast}\left[GW\right]$} & {\tiny{}0.098} & {\tiny{}$\Phi\left[GW\right]$} & {\tiny{}0.093}\tabularnewline
\hline 
{\tiny{}Toshiba (T)} & {\tiny{}-0.4569} & {\tiny{}0.8412} & {\tiny{}0.9575} & {\tiny{}$\alpha^{RC}\left[MW/K\right]$} & {\tiny{}3.0056} & {\tiny{}$P\left[GW\right]$} & {\tiny{}0.342} & {\tiny{}$P\left[GW\right]$} & {\tiny{}0.342} & {\tiny{}$P\left[GW\right]$} & {\tiny{}0.342}\tabularnewline
\cline{5-12} 
 &  &  &  & {\tiny{}$^{\gamma^{RC}}$} & {\tiny{}2.8077} & {\tiny{}$\eta$} & {\tiny{}0.48} & {\tiny{}$\eta^{\ast}$} & {\tiny{}0.630} & {\tiny{}$\eta$} & {\tiny{}0.63}\tabularnewline
\cline{5-12} 
 &  &  &  & {\tiny{}$\tau^{RC}$} & {\tiny{}0.2012} & {\tiny{}$\Phi\left[GW\right]$} & {\tiny{}0.23} & {\tiny{}$\Phi^{\ast}\left[GW\right]$} & {\tiny{}0.097} & {\tiny{}$\Phi\left[GW\right]$} & {\tiny{}0.092}\tabularnewline
\hline 
\end{tabular}
\par\end{centering}

\end{table}

The curves that can be generated from a point in the configuration
space and represent an operating regime for a thermal machine,  contain also other operating regimes that are achievable, as long as there 
are physically acceptable relationships for the control parameters.

\section{Conclusions}
\label{sec:4}
Although, the way thermal machines exchange heat with the surroundings depends strongly on the type of the proposed heat transfer law, 
their energy optimization goes beyond the irreversibilities that can be introduced in the model. The type of irreversibilities the CA model
does not take into account are the internal ones of the system (working
substance), that in practice are difficult to quantify. Nevertheless,
the CA model provides boundary operating conditions that take into
account both the energy (power output, efficiency and dissipation)
and the parameters related to the construction of the energy converter
(thermal conductance and temperatures quotient). On the other hand, the
CA model allows to establish the concept of objective function, a relationship
between the process variables that exhibits an optimal behavior.

Within the generalization proposals for the efficient power as an objective
function, the $k$--Efficient Power presents a clear advantage, it
can characterize modes of operation that are not achievable for the
other proposals. As the high reduced temperature ($a_{h}$) that characterizes
the maximum $k$--Efficient Power regime can be written in terms of
the generalization parameter $k$, a biunivocal relationship between
both variables is established. This means, it is possible to associate
any process variable with this generalization parameter and study
all the physically accessible operating regimes at the same time.

The conversion efficiency  usually is found among the reported data of power plants. 
For this reason, it is a candidate that permits to
find the value of $k$, since this function requires a minimum amount
of system information [$k(\eta,\tau)$]. This dimensionless parameter
allows us to say, in the particular case of power plants,
if their operation modes are within the HE zone or outside it. 
While in the case of power or dissipation, we require more information from
the system to locate the operation modes in the configuration space (HE or LE). This extra information must be estimated, 
because the control parameters $\alpha$ and $\gamma$ modulate the amount of useful work obtained.

With the $k$--Efficient Power model that we have been analyzing, we found
the most suitable configurations for $\alpha,\gamma$,
$\tau$, $P$ and $\eta$ that fit the operation of some power plants. Likewise, we show the parameter $k$ is capable
of building any energy curve ($P$ \textit{vs.} $\eta$ and $P$ \textit{vs.} $\Phi$) compatible
with some attainable regime. If we move the characteristic operating
point for each power plant in this set of configurations on
a curve with an optimal energetic performance, we can establish the ``restructuring
conditions'' on the control parameters for all plants that are operating
in the LE zone. The advantage of this new configuration is that the
plant will operate in the HE zone, decreasing its dissipation without
sacrificing power output. Such is the case of the monocycle plants
(see Fig. \ref{fig:MejoraEnEnergetica}), those plants are operating outside
the HE zone and also the Toshiba 109FA, which is a combined cycle. All of the above 
is summarized in the following: the
operating regimes, which come from an objective function and that are also associated to a generalization parameter determine in the space of energy configurations,
families of curves linked to the values of the control parameters
(design). If the values of these parameters are modified, the
optimal performance of an energy converter is guaranteed without having
to build a new one.

\section*{Acknowledgement}

Thanks to F. Angulo-Brown for his recommendations to improve the manuscript. 
This work was partially supported by CONACYT Grants: 288669 and 308401 
Instituto Polit\'ecnico Nacional: SIP-project number: 20181897,
COFAA-Grant: 5406, EDI-Grant: 1750 and SNI-CONACYT Grant: 16051, MEXICO.

\end{document}